\documentclass[11pt,titlepage,a4paper]{article}
\usepackage{Bozho-LaTeX-Style}	
\usepackage{Bozho-LaTeX-Logo}	
\usepackage{cite}	

\title{\bfseries	\vspace*{-1.678902345in}
{\huge Transformation laws of the\\[1ex]
components of classical and quantum fields\\[1ex]
and Heisenberg relations }
}

\vspace{1.7ex}

\author{
Bozhidar Z.\ Iliev
\thanks{Laboratory of Mathematical Modeling in Physics,
Institute for Nuclear Research and \mbox{Nuclear} Energy,
Bulgarian Academy of Sciences,
Boul.\ Tzarigradsko chauss\'ee~72, 1784 Sofia, Bulgaria}
\thanks{E-mail address: bozho@inrne.bas.bg}
\thanks{URL: http://theo.inrne.bas.bg/$\sim$bozho/}
}

\date{
 \vspace{2.27ex}\ShortTitle{Transformation laws of physical fields}	\\[0.27ex]
 \vspace{3.27ex}
\small
	\begin{tabular}{r@{$\colon\to~$}l}
	\end{tabular} \\[1.27ex]
 \small
	\begin{tabular}{r@{$\colon~$}l}
	\end{tabular} \\[-0.27ex]
\normalsize
 \vspace{4.27ex}{\Huge\BOZHO}	\\[4.27ex]
	\begin{tabular}{r@{\hspace{0.512em}}|@{\hspace{0.512em}}l}
\vspace{0.27ex}	
&
\vspace{0.27ex}
	\end{tabular} \\[1.27ex]
}

	\newcommand{\base}{\mathit{M}}	

\begin{document}		

\renewcommand{\thepage}{\roman{page}}

\renewcommand{\thefootnote}{\fnsymbol{footnote}} 
\maketitle				
\renewcommand{\thefootnote}{\arabic{footnote}}   

\tableofcontents		


\begin{abstract}

	The paper recalls and point to the origin of the transformation laws of
the components of classical and quantum fields. They are considered from the
"standard" and fibre bundle point of view. The results are applied to the
derivation of the Heisenberg relations in quite general setting, in particular,
in the fibre bundle approach. All conclusions are illustrated in a case of
transformations induced by the Poincar\'e group.

\end{abstract}

\renewcommand{\thepage}{\arabic{page}}


\section {Introduction}
\label{Introduction}

The components of physical fields with respect to a reference frame change when
this frame changes. The correspondence between the components of a field with
respect to two frames of reference is known as a transformation law of that
field. These laws are important characteristics of the physical fields; \eg via
them one can make a distinction between vector and spinor fields.

	This paper recalls and shows the origin of the transformation laws of
the physical fields in the classical case (section~\ref{Sect2}) and in the
quantum one (section~\ref{Sect3}). A  conclusion is made that in the
transformations of the classical fields are involved passive coordinate
transformations while in the quantum case are presented active coordinate
transformations, \ie diffeomorphisms associated with coordinate
transformations. In section~\ref{Sect4} are considered the transformation laws
of the fields from fibre bundle point of view in which the fields are described
as sections of suitable vector bundles. In this setting in the transformation
laws of classical as well as quantum fields are involved only passive
coordinate transformations.  Special attention is paid to transformations
induced by the Poincar\'e group.

	The results obtained are applied in section~\ref{Sect5} to
derivation of the so\ndash called Heisenberg relations in quite general
setting.~%
\footnote{~%
Section~\ref{Sect5} is an extended version of~\cite{bp-Heisenberg-rel}.%
}
 The investigation start withs transformations induced by the Poincar\'e group;
in particular, some well known results are reproduced. The consideration of
internal transformations leads to Heisenberg relations concerning different
charges. The Heisenberg relations are derived also in the general setting when
the transformations are induced by three representations of a given group. At
last, we look on the Heisenberg relations from fibre bundle view point.

	Section~\ref{Conclusion} closes the work with consideration of some
problems concerning the observability and measurability of the reference
frames and the components of the physical fields in them.


\section{Transformation laws of classical fields}
\label{Sect2}

	Let $\base$ be the Minkowski spacetime model of special relativity. A
(classical) field $\varphi$ describing (some property of) a physical system is
a mapping $\varphi\colon\base\to V$ where $V$ is a real or complex vector
space of finite dimension.~%
\footnote{~%
In a case of Newtonian physics $M$ must be the Newtonian 3-dimensional absolute
space.%
}$^{,~}$%
\footnote{~%
It seems there are not real classical systems described via infinite demensional
vector spaces. But in the fibre bundle quantum case infinitely dimensional
spaces may arise naturally.%
}
Usually $V$ is a space in which a representation
of the Poincar\'e group (in particular, its subgroup the Lorentz group) acts
(\emph{vide infra}). The field $\varphi$ is equivalently described via its
components $\varphi^i$, $i=1,\dots,\dim V$, in some frame $\{e_i\}$ ~%
\footnote{~%
That is $e_i \colon x\to e_i(x)\in V$ and $\{e_i(x)\}$ is a basis of $V$.%
}
over $\base$ in $V$, i.e.~%
\footnote{~%
From here on in our text the Latin indices $i,j,k,\dots$ run from 1 to the
dimension $\dim V$ of $V$ and the summation convention over indices repeated
on different levels is assumed over the whole range of their values.%
}
	\begin{equation}	\label{2.1}
\varphi^i\colon\base\to\field
\quad
\varphi(x) =: \varphi^i(x) e_i(x)
\qquad
x\in\base,
	\end{equation}
where $\field=\field[R],\field[C]$, depending on whether a real or complex field
is considered, and we write $\varphi=\varphi^i e_i$ meaning that $\varphi^i
e_i\colon x\mapsto\varphi^i(x) e_i (x)=\varphi(x)$. Often a field $\varphi$ is
depicted as a vector\ndash column $(\varphi^1,\dots,\varphi^{\dim V})^\top$,
$\top$ being the matrix transposition sign, formed from the \field\ndash valued
functions $\varphi^i$, called the \emph{components} of $\varphi$. One should be
aware of the fact that in the (physical) literature as components of $\varphi$
are referred the (local, if $\base$ is a general manifold) representations
	\begin{equation}	\label{2.2}
\varphi_{ u }^{i} := \varphi^i\circ u ^{-1}
	\colon \field[R]^4 \to \field
	\end{equation}
of $\varphi^i$ in some (local) chart $(U,u)$ of $\base$ with
$ u \colon \base\supseteq U\to\field[R]^4$. However, since as a set $\base$
coincides with  $\field[R]^4$ and one (implicitly) thinks of $ u $ as
the identity mapping of $\field[R]^4$, practically everywhere in the
literature the authors write $\varphi^i$ and talk about it having in mind and
dealing, in fact, with $\varphi_{ u }^{i}$.

	The transformation laws of fields (more precisely,
of their components) are quite important their characteristics. Before
formulating them explicitly, some remarks have to be made.

	Mathematically the components $\varphi^i$ of a field $\varphi$ are the
functions $\varphi^i\colon\base\to\field$ as defined above. They are defined
relatively to some frame $\{e_i\}$ over $\base$.
Similarly, the coordinates $x^\mu$, $\mu=0,1,2,3$,~%
\footnote{~%
In this paper the Greek indices  $\lambda,\mu,\nu,\dots$ run from 0 to
$3=\dim\base-1$ and refer to the Minkowski space $\base$.%
}
of a spacetime point $x\in\base$ are the numbers $x^\mu\in\field[R]$ defined
by the relation $ u (x)=:(x^0,x^1,x^2,x^3)=:\Mat{x}$.  They are given with
respect to a coordinate system
$\{u^\mu\}$ with $u^\mu \colon x\mapsto u^\mu(x)=:x^\mu $.

	Physically the components of a field $\varphi\colon\base\to V$ or the
coordinates of a point $x\in\base$ are defined with respect to a given
\emph{reference frame}. Among other things, the last concept includes a way of
measuring the spacetime coordinates of points and the components of all fields
concerning particular problem (system)~%
\footnote{~%
See section~\ref{Conclusion} for some details concerning measurability of
fields components.%
};
 in our case the functions $\varphi^i$
(or their values $\varphi^i(x)=\varphi_{ u }^{i}( u (x))\in\field$) and the
numbers $x^\mu$. So, in the particular situation, a frame of reference is
mathematically described via a pair $( u ,\{e_i\})$ of a coordinate system~%
\footnote{~%
As $u(x)=(u^0(x),\dots,u^3(x))\in\field[R]^4$, we identify the coordinate
homeomorphism $u \colon U\to \field[R]^4$ with
 $(u^0,\dots,u^3) \colon x\mapsto (u^0(x),\dots,u^3(x))$ and, using some
freedom of the language, we call $u$ also coordinate system regardless that the
last term means $\{u^\mu\}$ or $(u^0,\dots,u^3)$. Note also that
$u^\mu=r^\mu\circ u$ where $\{r^\mu\}$ is the standard Cartesian coordinates
system on $\field[R]^4$, \ie  $r^\mu \colon (x^0,\dots,x^3)\to x^\mu$.%
}
$u$ of $\base$ and frame $\{e_i\}$ over $\base$ in $V$. We write $u$ at first
position in $(u,\{e_i\})$ as generally the frames $\{e_i\}$ may depend on $u$
(see below subsection~\ref{Subsect3.2}).

	Suppose $( u ,\{e_i\})$ and $( u ',\{e'_i\})$ represent
two reference frames and
    \begin{equation}    \label{2.2-1}
e'_i(x)=A_i^j(x)e_j(x)
    \end{equation}
for some non\ndash degenerate matrix\ndash valued function $A=[A_i^j]$. We
admit that $e_i$ (resp.\ $e'_i$) may depend on $ u $ (resp.\ $ u '$), so $A$
may depend on $ u $ and $ u '$.~%
\footnote{~%
More precisely, $A$ may depend on the transformation
$ u \mapsto u '$.%
}
A \emph{transformation law} of a field $\varphi$ (relative to
$( u ,\{e_i\})$ and $( u ',\{e'_i\})$) is called the
correspondence
$\varphi_{ u }^i\mapsto\varphi_{ u '}^{\prime\,i}$.
Writing the decompositions
$ \varphi(x) = \varphi^i(x) e_i(x) =  \varphi^{\prime\,i}(x) e'_i(x) $
and using
 $\varphi_{ u }^i:=\varphi^i\circ u ^{-1}$,
\(
\varphi_{ u '}^{\prime\,i}
	:=\varphi^{\prime\,i}\circ u ^{\prime\,-1}
\)
and $e'_i=A_i^je_j$, we get the explicit form of the transformed field
components (relative to $( u ',\{e'_i\})$):
	\begin{align}    \label{2.3}
\varphi^{\prime\, i}(x)
&= (A^{-1}(x))_j^i \varphi^j(x)
\\ 			\tag{\ref{2.3}$^\prime$}
			\label{2.3'}
\varphi_{ u '}^{\prime\,i} (r)
&=
\bigl( A^{-1}( u ^{\prime\,-1}(r)) \bigr)_{j}^{i}  \,
	\varphi_ u ^j\bigl(
	( u \circ u ^{\prime\,-1}) (r)
	\bigr)
	\end{align}
for any $x\in U\cap U'\subseteq \base$ and $r\in u(U)\cap
u'(U')\subseteq\field[R]^4$.

	Usually, in the physical literature, this formula is written in the
following more concrete form. Let $ u $ and $ u '$ be \emph{linear} and $ u '$
be obtained from $ u $ by a Poincar\'e transformation, \viz
    \begin{equation}    \label{2.3-2}
u '(x)= \Lambda u(x)+a
    \end{equation}
with fixed $a\in\field[R]^4$ and $\Lambda$ being a Lorentz transformation (\ie
the matrix of a 4\ndash rotation). Let in $V$ acts a representation $D$ of the
Lorentz (Poincar\'e) group under which $\varphi$ transforms, \ie $ u \mapsto u
'$ implies
    \begin{equation}    \label{2.3-3}
A^{-1}(x)=\Mat{D}(\Lambda,a)
    \end{equation}
for all $x\in\base$ with $\Mat{D}(\Lambda,a)=[D_j^i(\Lambda,a)]$ being the
matrix of $D(\Lambda,a)\colon V\to V$, corresponding to~\eref{2.3-2}, in
$\{e_i\}$. Then equation~\eref{2.3'} reduces to
    \begin{equation}    \label{2.3-1}
\varphi_{u'}^{\prime\, i}(r)
= {D}_j^i(\Lambda,a) \varphi_u^j(\Lambda^{-1}(r-a)) ,
    \end{equation}
where we have used the first of the following simple corollaries
from~\eref{2.3-2}:
    \begin{subequations}    \label{2.3-4}
    \begin{align}	    \label{2.3-4a}
(u\circ u'^{-1}) (r) &= \Lambda^{-1}(r-a)
\\			    \label{2.3-4b}
(u'\circ u^{-1}) (r) &= \Lambda r + a
\\			    \label{2.3-4c}
(u^{-1}\circ u') (x) &= u^{-1}(\Lambda u(x) + a)
\\			    \label{2.3-4d}
(u^{\prime\, -1}\circ u) (x) &= u^{-1}(\Lambda^{-1}(u(x)-a)).
    \end{align}
    \end{subequations}

	If we identify $x\in\base$ with $ u (x)=(x^0,\dots,x^3)\in\field[R]^4$
and omit the indices $ u $ and $ u '$, equation~\eref{2.3-1} takes the
familiar form~\cite[\S~2]{Rumer&Fet}
	\begin{equation}	\label{2.4}
\varphi^{\prime\,i} (x)
=
{D}_j^i(\Lambda,a) \varphi^j(\Lambda^{-1}(x-a)).
	\end{equation}


\section {Transformation laws of quantum fields}
\label{Sect3}

	After quantization a classical field $\varphi$ transforms into a
vector operator\ndash valued distribution (generalized function)
\(
\Mat{\varphi}
=
\sum_{i=1}^{\dim V} (0,\dots,0,\Mat{\varphi}_i,0,\dots,0),
\)
where $\Mat{\varphi}_i$ is usual operator\ndash
valued distribution which sits into the $i^{\text{th}}$ position. The action
of $\Mat{\varphi}$ on a vectorial test function
$\Mat{f}=(f^1,\dots,f^{\dim V})$,  $f^i \colon \field[R]^4\to\field$, is
often written symbolically as
	\begin{equation}	\label{2.5}
\Mat{\varphi}(\Mat{f})
=:
\int\sum_{i} \varphi_i(r) f^i(r) \Id^4r
	\end{equation}
where the integration is over $\field[R]^4$ and $\varphi_i(r)$ are operators
in the system's Hilbert space of states that are treated as operator\ndash
valued components of the (nonsmeared) quantum field.~%
\footnote{~%
For a rigorous description of quantum fields,
see~\cite{Bogolyubov&et_al.-AxQFT,Bogolyubov&et_al.-QFT}.%
}

	Let $O$ and $O'$ be two observers (reference frames). Since the
functional $\Mat{\varphi}$ describes a quantum field irrespectively of any
observers, the transition $O\mapsto O'$ should imply
$\Mat{\varphi}\mapsto\Mat{\varphi}$ but the functions on which
$\Mat{\varphi}$ acts can be observer\ndash dependent, \ie $\Mat{f}\mapsto \Mat{f}'$,
where $\Mat{f}'$ (or, more precisely, $\Mat{\varphi}(\Mat{f}')$) may be determined as
follows. Suppose $U$ is an (invertible, linear, and, possibly, unitary)
operator representing the change $O\mapsto O'$ of the state vectors of the
system (the fields, in our case)~%
\footnote{~%
We use one and the same letter $U$ to denote the just mentioned operator and a
neighborhood in the Minkowski spacetime and hope that this will not lead to
misunderstandings further in this paper.%
},
\ie if $X$ is a state vector relative to
$O$, the vector $U(X)=X'$ represents the same state with respect to  $O'$. As
$\Mat{\varphi}(\Mat{f}')$ plays with respect to $O'$ the same role as
$\Mat{\varphi}(\Mat{f})$ relative to $O$, the vector
$U((\Mat{\varphi}(\Mat{f}))(X))$, representing $(\Mat{\varphi}(\Mat{f}))(X)$
relative to $O'$, and $(\Mat{\varphi}(\Mat{f}'))(U(X))$, representing the
action of $\Mat{\varphi}(\Mat{f}')$ on the transformed vector $U(X)$, should be
equal, $U((\Mat{\varphi}(\Mat{f}))(X)) = (\Mat{\varphi}(\Mat{f}'))(U(X))$.
Therefore, we have~%
\footnote{~%
The above discussion is not rigorous and should be  considered only as a
motivation. In fact, equation~\eref{2.6} below must be postulated. Similarly,
equation~\eref{2.8-2} can be considered as a definition of $\varphi'_i$.%
}
	\begin{equation}	\label{2.6}
\Mat{\varphi}(\Mat{f}') = U\circ\Mat{\varphi}(\Mat{f})\circ U^{-1} .
	\end{equation}

	The transition $\Mat{\varphi}(\Mat{f})\mapsto\Mat{\varphi}(\Mat{f}')$
can effectively and equivalently be describe by admitting that the test
function $f$ remains unchanged, while the field $\Mat{\varphi}$ transforms into
$\Mat{\varphi}'$, \ie
	\begin{equation}	\label{2.8}
\Mat{\varphi}'(f) = \Mat{\varphi} (f') .
	\end{equation}
From here and~\eref{2.6} one immediately gets
    \begin{equation}    \label{2.8-1}
\Mat{\varphi}' = U\circ \Mat{\varphi}\circ U^{-1}
    \end{equation}
or equivalently
    \begin{equation}    \label{2.8-2}
\varphi'_i(r) = U\circ \varphi_i(r)\circ U^{-1} .
    \end{equation}

	If we have three observers $O_1$, $O_2$ and $O_3$ and $U_{ab}$,
$a,b=1,2,3$, maps the state vectors relative to $O_b$ into ones relative to
$O_a$, then
    \begin{align}    \label{2.6-1}
& U_{aa} = \id
\\		    \label{2.6-2}
& U_{ab}\circ U_{bc} = U_{ac} ,
    \end{align}
where $\id$ is the identity mapping, and hence the set
$\{U \colon U$ maps state vectors between two observers$\}$
 has a structure of a ``partial'' monoid (groupoid with identity/unit element
whose operation (multiplication) is not defined on all elements). In a case of
inertial observers (connected via Poincar\'e transformations) the mappings
$U$ form a representation of the Poincar\'e group and therefore $\{U\}$ has a
group structure.

	Now we want to find the analogue of~\eref{2.3'} (in
particular, of~\eref{2.4}) in the quantum case, \ie we would like to derive
the transformation law of the \emph{components} of a quantum field
$\Mat{\varphi}$.

	Treating $f^j$ as components of a classical field and writing $f^j_u$
for $f^j$, we, due to~\eref{2.3'}, have
\begin{equation}	\label{2.9-1}
f_{u'}^{\prime\,i}(r)
=
\bigl( A^{-1}( u'^{-1}(r)) \bigr)_j^i f_u^j( ( u \circ u'^{-1})(r) )
	\end{equation}
which in a case of Poincar\'e transformation reduces to
	\begin{equation}	\label{2.9-2}
f_{u'}^{\prime\,i} (r)
=
	{D}_j^i(\Lambda,a)
	f_u^j ( \Lambda^{-1}(r-a) ) .
	\end{equation}
If we identify $x$ and $r=u(x)$ and omit the subscripts $u$ and $u'$,
then the last equation reads
	\begin{equation}	\label{2.9-3}
f_{}^{\prime\,i}(x)
=
	{D}_j^i(\Lambda,a)
	f^j ( \Lambda^{-1}(x-a) )
	\end{equation}
which coincides with~\cite[p.~249, eq.~9.5]{Bogolyubov&et_al.-AxQFT} up to
notation.

	The above-said implies that $\Mat{\varphi}$ (resp.\ $\Mat{f}$) can be
regarded as observer independent (resp.\ dependent), \ie with respect to
any observer $O$ its components $\varphi_i(r)$ (resp.\ $f_{ u }^{i}(r)$)
are independent of (resp.\ dependent on) the coordinate system $ u $ and frame
$\{e_i\}$ associated with $O$ (resp.\ and transforming according to~\eref{2.3'}
when $O$ is replace with $O'$). So, we have   ~%
\footnote{~%
For the correctness of the integrals (over $\field[R]^4$) below, \eg
in~\eref{2.7}, we have to assume that the neighborhoods $U$ and $U'$ are such
that $u(U\cap U')=u'(U'\cap U)=\field[R]^4$ which is equivalent to
$u(U)=u'(U')=\field[R]^4$. However, this assumption is not needed for the
equations in which integrals are not involved in which case we require $r\in
u(U\cap U')\cap u'(U'\cap U)$.%
}
	\begin{equation}	\label{2.7}
	\begin{split}
\Mat{\varphi}(\Mat{f}')
& =
\int\varphi_i(r) f_{ u '}^{\prime\,i}(r) \Id^4r
=
\int\varphi_i(r) \bigl( A^{-1}( u ^{\prime\,-1}(r)) \bigr)_j^i
 	f_{ u }^{j}
	  ( ( u \circ u ^{\prime\,-1}) (r) ) \Id^4r
\\
& =
\int\varphi_i( ( u '\circ u ^{-1}) (q) )
	\bigl( A^{-1}( u ^{-1}(q)) \bigr)_j^i f^j_{ u }(q)
	\frac{\pd( u' \circ u ^{-1})(q)}{\pd q} \Id^4 q ,
	\end{split}
	\end{equation}
where~\eref{2.3'} was used, the variable $r$ has been changed to
$q=( u \circ u ^{\prime\,-1})(r)$,
and
$\frac{\pd( u '\circ u ^{-1})(q)}{\pd q} $
is (the symbolic notation for) the Jacobian of the change $r\mapsto q$.

	Now, writing
 $\Mat{\varphi}'(\Mat{f})=\int\sum_i\varphi'_{u',i}(q) f_ u ^i (q) \Id^4q$
(see~\eref{2.5}), from~\eref{2.8} and~\eref{2.7}, we get
	\begin{equation}	\label{2.9}
\varphi'_{u',i}(r)
=
\frac{\pd( u '\circ u ^{-1})(r)}{\pd r}
	\bigl( A^{-1}( u ^{-1}(r)) \bigr)_i^j
	\varphi_{u,j} ((  u '\circ u ^{-1})(r) ) .
	\end{equation}

	In particular, if $ u $ and $ u '$ are linear, $O$ and $O'$ are
connected via (proper) Poincar\'e transformation,
$ u '(x)= \Lambda u(x)+a$ with $x\in\base$, then~\eref{2.9} reduces to
    \begin{equation}    \label{2.9-4}
\varphi'_{u',i}(r) = {D}_i^j(\Lambda,a) \varphi_{u,j} (\Lambda r+a)
    \end{equation}
due to~\eref{2.3-4b}. Here we have used that the Jacobian of
$ u \mapsto u'$ equals to one (for proper transformations) and
$\Mat{D}(\Lambda,a)$ is the same matrix which appears in~\eref{2.4}.
	If $r=u (x)$ is identified with $x$ and the subscripts $u$ and $u'$
are omitted, this formula reduces to the usual one
(see, e.g.,~\cite[\S~10]{Rumer&Fet} and \cite[\S~68]{Bjorken&Drell-2})
	\begin{equation}	\label{2.10}
\varphi'_i(x) = {D}_i^j(\Lambda,a) \varphi_j(\Lambda x+a) .
	\end{equation}
Notice, the classical fields $\varphi^i$ transform according to~\eref{2.4}
with $\Mat{D}(\Lambda,a)$, while the quantum ones transform via~\eref{2.10} in
which the transposed matrix $\Mat{D}^\top(\Lambda,a)$ of $\Mat{D}(\Lambda,a)$ is
utilized.

	The above discussion can be summarized in the following three equations
    \begin{align}    \label{2.17}
& \varphi'_i(r) = U\circ \varphi_i(r)\circ U^{-1}
\\		    \label{2.18}
& \varphi'_{u',i}(r)
=
\frac{\pd( u '\circ u ^{-1})(r)}{\pd r}
	\bigl( A^{-1}( u ^{-1}(r)) \bigr)_i^j
	\varphi_{u,j} ((  u '\circ u ^{-1})(r) )
\\		    \label{2.19}
& U\circ\varphi_{u,i}(r)\circ U^{-1}
=
\frac{\pd( u '\circ u ^{-1})(r)}{\pd r}
	\bigl( A^{-1}( u ^{-1}(r)) \bigr)_i^j
	\varphi_{u,j}( ( u '\circ u ^{-1})(r) )
    \end{align}
the last of which is a consequence of the preceding two ones. At this point we
have to say that equation~\eref{2.17} is a consequence of the
hypotheses/assumptions~\eref{2.6} and~\eref{2.8}, while~\eref{2.18} is a result
of the hypotheses/assumptions~\eref{2.8} and~\eref{2.9-1}. If all of the
equations~\eref{2.6}, \eref{2.8} and~\eref{2.9-1} hold, we can write the chain
equality
	\begin{equation}	\label{2.11}
\varphi'_{u',i}(r)
=
U\circ\varphi_{u,i}(r)\circ U^{-1}
=
\frac{\pd( u '\circ u ^{-1})(r)}{\pd r}
	\bigl( A^{-1}( u ^{-1}(r)) \bigr)_i^j
	\varphi_{u,j}( ( u '\circ u ^{-1})(r) )
	\end{equation}
which, in the above special case of Poincar\'e transformation, reduces to the
known result~\cite[eq,~(11.67)]{Bjorken&Drell-2}
	\begin{equation}	\label{2.12}
\varphi'_i(x)
=
U(\Lambda,a)\circ\varphi_i(x)\circ U^{-1}(\Lambda,a)
=
{D}_i^j(\Lambda,a)\varphi_j(\Lambda x+a) ,
	\end{equation}
where $x$ is identified with $u(x)$, the subscripts $u$ and $u'$ are omitted,
and  $U(\Lambda,a)$ is the element corresponding to~\eref{2.3-2} via a
representation of the Poincar\'e group on the space of state vectors.  ~%
\footnote{~%
The equation~\eref{2.12}, which holds in relativistic quantum mechanics too,
can be derived also under the following assumptions
(cf.~\cite[\S68]{Bjorken&Drell-2}):
	(i)~the state vectors do not change under a (passive) Poincar\'e
transformations, $X(x)=X(u^{-1}(r))=X(u'^{-1}(r'))$ for $r=u(x)$, $r'=u'(x)$
and $r'=\Lambda r + a$;
	(ii)~there is a unitary operator $U(\Lambda,a)$ such that
$X'(x)=U(\Lambda,a)(X(x))$;
	(iii)~there is a non-degenerate matrix
$D(\Lambda,a)=[D_i^j(\Lambda,a)]$ such that
\(
  \varphi_i\circ u'^{-1}(r)
= \varphi'_{u',i}(r')
= D_i^j(\Lambda,a) \varphi_j( ( u'\circ u^{-1} )(r) )
\):
	(iv)~the scalar products in system's Hilbert space of states are
invariant, \ie
\(
\langle X'(u'^{-1}(r')) | \varphi'_{u',i}(r')(Y'(u'^{-1}(r')) \rangle
=
\langle X(u^{-1}(r)) | \varphi_{u,i}(r)(Y(u^{-1}(r)) \rangle
\).
	In fact, these suppositions immediately imply
\(
U^{-1}(\Lambda,a)\circ D_i^j(\Lambda,a)\varphi_{u,j}(r) \circ U(\Lambda,a)
= \varphi_{u,i}(r),
\)
\ie
\(
U^{-1}(\Lambda,a)\circ \varphi_{u,i}(r) \circ U(\Lambda,a)
=  D_i^j(\Lambda,a) \varphi_{u,j}(r)
=\varphi'_{u',i}(r).
\)
Now, identifying $x$ with $r=u(x)$ and writing $\varphi_i$ for $\varphi_{u,i}$
and $\varphi'_i$ for $\varphi'_{u',i}$, we obtain~\eref{2.12}.%
}

	If we set
    \begin{equation}    \label{2.12-1}
\varphi_{u,i}=\varphi_i\circ u^{-1}
\qquad
f_u^i=f^i\circ u^{-1}
    \end{equation}
and $x=u^{\prime\, -1}(r)\in\base$, then we can rewrite~\eref{2.9-1}
and~\eref{2.11} respectively as
    \begin{align}    \label{2.13}
f^{\prime\, i}(x)
&= (A^{-1}(x))^i_j f^j(x)
\\		    \label{2.14}
    \begin{split}
\varphi'_i(x)
&=
U\circ \varphi_i((u^{-1}\circ u')(x)) \circ U^{-1}
\\
&=
\frac{\pd(u'\circ u^{-1})(r)}{\pd r} \Big|_{r=u'(x)}
\bigl( A^{-1}( (u^{-1}\circ u') (x)) \bigr)_i^j
\varphi_j( (u^{-1}\circ u')^2 (x) ).
    \end{split}
    \end{align}
	In the special case of a Poincar\'e transformation, the last two
equations reduce to (see~\eref{2.3-3} and~\eref{2.3-4b})
    \begin{align}    \label{2.15}
f^{\prime\, i} (x)
& = {D}_j^i(\Lambda,a) f^j(x)
\\		    \label{2.16}
    \begin{split}
\varphi'_i(x)
& =
U(\Lambda,a)\circ \varphi_i(u^{-1}(\Lambda u(x) + a))\circ U^{-1}(\Lambda,a)
\\
&=
{D}_i^j(\Lambda,a) \varphi_j (u^{-1}(\Lambda(\Lambda u(x) + a) + a) ) .
    \end{split}
    \end{align}

	In this way we see a very essential difference between the
transformation laws of classical fields, like $f^i$, and quantum once, like
$\varphi_i$, when the frame of reference is changed:
    \begin{itemize}
\item
The classical fields transform according to \emph{passive} coordinate
transformations, \ie in the both sides of~\eref{2.13} (or~\eref{2.9-1}) are
involved quantities evaluated \emph{at one and the same spacetime point} $x$.

\item
Contrary to the previous observation, the quantum fields transform according
to \emph{active} coordinate transformations which means that in the both sides
of~\eref{2.14} (cf.~\eref{2.9}) are involved quantities evaluated
\emph{at different spacetime points}, viz.\ $x$, $(u^{-1}\circ u')(x)$ and
$(u^{-1}\circ u')^2(x)$, the last two of which are obtained from $x$ via the
(local) diffeomorphism
$u^{-1}\circ u'  \colon \base\supseteq U\cap U'\to U\cap U'$ which
in turn is one of the two possible active interpretations of the change
$u\mapsto u'$.~%
\footnote{~%
The another diffeomorphism is
$u^{\prime\,-1}\circ u \colon \base\supseteq U\to U'\subseteq\base$ , which is
the inverse of $u^{-1}\circ u'$ on $U\cap U'$.%
}
    \end{itemize}


\section {Bundle view on the transformations laws of fields}
\label{Sect4}

	The physical fields, classical and quantum ones, can be represented as
sections of vector bundles. Such a view-point brings  an additional light on
their transformation properties.

\subsection {Changes of frames in the bundle space}
\label{Subsect4.1}

	Suppose a physical field is described as a section
$\varphi \colon \base\to E$ of a vector bundle $(E,\pi,\base)$. Here $\base$ is
a real differentiable (4\ndash)manifold (of class at least $C^1$), serving as a
spacetime model, $E$ is the bundle space and $\pi \colon \base\to E$ is the
projection; the fibres $\pi^{-1}(x)$, $x\in\base$, are isomorphic vector
spaces.~%
\footnote{~%
To make a contact with section~\ref{Sect2}, one should identify $V$ with the
(standard) fibre of $(E,\pi,\base)$ and consider $\varphi(x)$ as an element of
$\pi^{-1}(x)$ rather then of $V$. Besides, now the Latin indexes refer to the
bundle space and run from~1 to the fibre dimension of the bundle.%
}
	Let $(U,u)$ be a chart of $\base$ and $\{e_i\}$ be a (vector) frame in
the bundle with domain containing $U$, \ie
 $e_i \colon x\mapsto e_i(x)\in\pi^{-1}(x)$ with $x$ in the domain of $\{e_i\}$
and $\{e_i(x)\}$  being a basis in $\pi^{-1}(x)$. Below we assume
$x\in U\subseteq \base$. Thus, similarly to~\eref{2.1}, we have
    \begin{equation}    \label{3.1}
\varphi \colon \base\ni x\mapsto \varphi(x)
=
\varphi^i(x) e_i(x)
=
\varphi_u^i(\Mat{x}) e_i(u^{-1}(\Mat{x})) ,
    \end{equation}
where
    \begin{equation}    \label{3.2}
\Mat{x}:=u(x) \qquad
\varphi_u^i:=\varphi^i\circ u^{-1}
    \end{equation}
and $\varphi^i(x)$ are the components of the vector
$\varphi(x)\in\pi^{-1}(x)$ relative to the basis $\{e_i(x)\}$ in $\pi^{-1}(x)$.

	Similarly, if $(U',u')$ and $\{e'_i\}$ are other chart and frame,
respectively, and $x\in U'$, then
    \begin{equation}    \label{3.3}
\varphi \colon \base\ni x\mapsto \varphi(x)
=
\varphi^{\prime\, i}(x) e'_i(x)
=
\varphi_u^{\prime\, i}(\Mat{x}') e'_i(u^{-1}(\Mat{x}')) ,
    \end{equation}
where
    \begin{equation}    \label{3.4}
\Mat{x}':=u'(x) \qquad
\varphi_{u'}^{\prime\, i} := \varphi^{\prime\, i}\circ u^{\prime\, -1} .
    \end{equation}
Further we shall suppose that $x\in U\cap U'\not=\varnothing$.

	We can write the expressions
    \begin{align}    \label{3.5}
e'_i(x) &= A_i^j(x) e_j(x)
\\
\tag{\ref{3.5}$'$}		    \label{3.5'}
e_i(x) &= (A^{-1}(x))_i^j e'_j(x)
    \end{align}
where $A \colon U\cap U'\to \GL(\dim\pi^{-1}(x),\field[R])$, $\GL(n,\field)$
being the general linear group of $n\times n$, $n\in\field[N]$, matrices over
$\field$, is the matrix-valued function defining the change
$\{e_i\}\mapsto\{e'_i\}$. Combining these expressions with~\eref{3.1}
and~\eref{3.3}, we get
    \begin{align}    \label{3.6}
\varphi_u^i(\Mat{x}) &= A_j^i(x) \varphi_{u'}^{\prime\, j}(\Mat{x}')
\\
\tag{\ref{3.6}$'$}		    \label{3.6'}
\varphi_{u'}^{\prime\, i}(\Mat{x}') &= (A^{-1}(x))_j^i \varphi_u^j(\Mat{x}) .
    \end{align}

	Generally the matrix $A(x)$ (resp.\ $A^{-1}(x)$) depends on the frames
$\{e_i\}$ and $\{e'_i\}$ and describes the transformation
$\{e_i\} \mapsto \{e'_i\}$ (resp.\ $\{e'_i\} \mapsto \{e_i\}$).  One may
reflect this by writing $A(\{e_i\}, \{e'_i\};x)$ instead of $A(x)$.~%
\footnote{~%
If $\{e''_i\}$ is a third frame, the following relations hold:
    \begin{align*}
A(\{e_i\}, \{e'_i\};x) A(\{e'_i\}, \{e''_i\};x)
& = A(\{e_i\}, \{e''_i\};x)
\\
A(\{e_i\}, \{e'_i\};x)
&= A^{-1}(\{e'_i\}, \{e_i\};x)
\\
A(\{e_i\}, \{e_i\};x)
&= \openone
    \end{align*}
with $\openone$ being the identity matrix. They express the simple fact that
the transformations between bases or frames form a representation of the
general linear group.%
}
At this point we can make a connection with physics, which will make the
above considerations more specific.

\subsection{Transformations induced by coordinate changes}
\label{Subsect3.2}

	The physical concept of a \emph{frame of reference} (or \emph{reference
frame}) is quite complex. However, for the purposes of the present work, it
reduces to a collection (ordered pair) $(u,e)$ of a coordinates system
$u=(u^0,u^1,u^2,u^3)$, associated with a chart $(U,u)$~%
\footnote{~%
We identify $u \colon U\to \field[R]^{\dim\base}$ with
\(
(u^0,\dots,u^{\dim\base}) \colon U\ni x\mapsto
(u^0(x),\dots,u^{\dim\base}(x)) \equiv u(x)$.%
}
and a frame $e=\{e_i\}$ in $E$ with domain containing $U$. The knowlage of
$(u,e)$ gives us the possibility to determine the coordinates of spacetime
points (via $u$) and the components of the fields (via $e$).

	It is a general opinion that the frames (and coordinates) in the bundle
space $E$ are not directly accessible for physical measurements.~%
\footnote{~%
The only known possible exception being  known (in a slightly modified form) as
the Aharonov-Bohm effect~\cite{Aharonov&Bohm,Bernstein&Phillips,Baez&Muniain},
but there are some doubts in its reality.%
}
For this reason it is accepted that to any coordinate system $u$ (or a chart
$(U,u)$) there corresponds a unique  frame $\{e_i\}$ in $E$ in which the
components of the physical fields $\varphi$ are determined.
	The mapping $u\mapsto e$ seems to be unknown in the general case with
 exception of the tangent bundle case, $E=T(M)$, in which it is assumed
$\{u^\mu\}\mapsto\{\frac{\pd }{\pd u^\mu}\}$, as well as its generalization to
arbitrary tensor bundles over $\base$. This is confirmed by the opinion that a
change $u\mapsto u'$ implies transforation $e\mapsto e'$ when one changes the
frame of reference from $(u,e)$ to $(u',e')$. Thus, if $(u,e)$ is a reference
frame (regardless of is $e$ induced by $u$) and we make a change
$(u,e)\mapsto(u',e')$, then $u'$ can be arbitrary (admissible) coordinate
system and the change $u\mapsto u'$ completely determines the transformation
$e\mapsto e'$.

	As a consequence of the above (non-rigorous) motivation, we assume
that, if $(u,e)$ is a reference frame, then the transformation $u\mapsto u'$
implies a change $(u,e)\mapsto(u',e')$. where $e'$ is given via~\eref{3.5} with
    \begin{equation}    \label{3.7}
A^{-1}(x)
\equiv A^{-1}(e,e';x)
= I(u\mapsto u'; x)
= I^{-1}(u'\mapsto u; x)
:= (I(u'\mapsto u; x))^{-1}
    \end{equation}
which means that
    \begin{align}    \label{3.8}
e'_i(x)
&= (I^{-1}(u\mapsto u'; x))_i^j e_j(x)
= I_i^j(u'\mapsto u; x) e_j(x)
\\
\tag{\ref{3.8}$'$}		    \label{3.8'}
e_i(x) &= I_i^j(u\mapsto u'; x) e'_j(x) ,
    \end{align}
where the matrix-valued function
$I(u\mapsto u'; \,\cdot\,) \colon U\cup U'\to \GL(\dim\pi^{-1}(x),\field[R])$
depends only on the change $u\mapsto u'$.
	Respectively now~\eref{3.6} and~\eref{3.6'} read:
    \begin{align}    \label{3.9}
\varphi_u^i(\Mat{x})
&= (I^{-1}(u\mapsto u'; x))_j^i \varphi_{u'}^{\prime\, j}(\Mat{x})
\\ \tag{\ref{3.9}$'$}		    \label{3.9'}
\varphi_{u'}^{\prime\, i}(\Mat{x}')
&= I_j^i(u\mapsto u'; x) \varphi_u^j(\Mat{x})
.
    \end{align}
If the (admissible) transformations $u\mapsto u'$ form (a representation of) a
group $G$, then the matrix-valued functions $I$ form a representation of $G$.

\subsection{Some peculiarities of the quantum fields}
\label{Subsect3.3}

		From fibre bundle point of view, the collection $\{\varphi_i\}$
of the (non smeared) components of a quantum field $\Mat{\varphi}$ in a given
reference frame is regarded as the set of the components of a section $\varphi$
of certain vector fibre bundle $(E,\pi,\base)$ over a manifold $\base$
identified with the Minkowski spacetime. This means that in a frame $\{e^i\}$
in $E$ over $\base$~%
\footnote{~%
For technical reasons we label the basic vector fields with superscripts, \ie
we write $e^i$ instead of $e_i$; respectively, below the quantum fields are
labeled via subscripts.%
}%
, we have
    \begin{equation}    \label{3.10}
\Sec(E,\pi,\base)\ni\varphi \colon \base\ni x\mapsto
\varphi(x) = \varphi_i(x) e^i(x) \in\pi^{-1}(x) .
    \end{equation}

	In this setting, the components $f^i$ of a vectorial test function
$\Mat{f}$ should also be regarded as components of a section $f$ of some vector
bundle $(F,\pi_F,\base)$ such that in some frame $\{e_i\}$ in $F$ over $\base$
is fulfilled
    \begin{equation}    \label{3.11}
\Sec(F,\pi_F,\base)\ni f \colon \base\ni x\mapsto
f(x) = f^i(x) e_i(x) \in\pi^{-1}_F(x) .
    \end{equation}

	To retain the validity of~\eref{2.5}, we should assume that
$(E,\pi,\base)$ is the bundle dual to $(F,\pi_F,\base)$,
$(F,\pi_F,\base)^*=(E,\pi,\base)$. Hance the sections of $(E,\pi,\base)$ are,
in fact, operator-valued linear mappings on the sections of $(F,\pi_F,\base)$.
Therefore, assuming $\{e^i\}$ to be the frame dual to $\{e_i\}$, i.e.\
$e^i=(e_i)^*$ with $e^i(e_j)=\delta_j^i$ ($=0$ for $i\not=j$ and $=1$ for
$i=j$), we obtain
    \begin{equation}    \label{3.12}
\varphi \colon f\mapsto
\varphi(f) \colon \base\ni x\mapsto
(\varphi(f))(x) :=
\varphi(x)(f(x)) = \sum_i\varphi_i(x)( f^i(x))
    \end{equation}
as a result of which equation~\eref{2.5} takes the form
	\begin{equation*}
\Mat{\varphi}(\Mat{f})
=
\int\varphi(f) .
	\end{equation*}
To give a rigorous meaning of this integral one needs the notion of
integration on manifolds (see, for instance,~\cite[ch.~IV]{Bruhat},
\cite[chapters~IV and~VII]{Greub&et_al.-1} and~\cite[ch.~VIII]{Lang/manifolds}).
To bypass this point, we shall assume below that the charts $(U,u)$ and
$(U',u')$ are such that $u(U)=u'(U')=\field[R]^4$ and the Jacobian of the
change $u\mapsto u'$ to be equal to one,
 $\frac{\pd (u'\circ u^{-1})(r)}{\pd r}=1$.
	Due to these assumptions, we can set
	\begin{equation*}
\Mat{\varphi}(\Mat{f})
=
\int_{\field[R]^4} (\varphi(f) \circ u^{-1})(r) \Id ^4r
	\end{equation*}
which expression is independent of $u$; to prove the last statement one may
write the last formula with $u'$ for $u$ and change the integration variable to
$q=(u\circ u^{\prime\, -1})(r)$.

	So, if we make the change
    \begin{subequations}    \label{3.13}
    \begin{align}    \label{3.13a}
& e^i\mapsto e^{\prime\, i} = A_j^i e^j
\\ \intertext{with a non-degenerate matrix-valued fucntion $A=[A_i^j]$, then
(cf.~\eref{3.5})}
		    \label{3.13b}
& \varphi_i(x)\mapsto \varphi'_i(x) = (A^{-1})_i^j(x) \varphi_j(x)
\\		    \label{3.13c}
& f^i(x)\mapsto f^{\prime\, i}(x) = A_j^j(x) f^j(x)
\\		    \label{3.13d}
& e_i(x)\mapsto e'_i(x) = (A^{-1})_i^j(x) e_j(x) .
    \end{align}
    \end{subequations}
If $(u,\{e^j\})$, $u=(u^0,\dots,u^3)$ being a coordinate system, is a
reference frame and we make a change
$(u,\{e^j\})\mapsto(u',\{e^{\prime\, j}\})$, then the components
    \begin{equation}    \label{3.13-1}
\varphi_{u,i}:=\varphi_i\circ u^{-1}
\qquad
f_u^i:=f^i\circ u^{-1}
    \end{equation}
of respectively $\varphi$ and $f$ transform into (cf.~\eref{3.6'})
    \begin{subequations}    \label{3.14}
    \begin{align}    \label{3.14a}
& \varphi'_{u',i}(\Mat{x}') = (A^{-1}(x))_i^j \varphi_{u,j}(\Mat{x})
\\		    \label{3.14b}
& f_{u'}^{\prime\,i}(\Mat{x}') = A_j^i(x) f_u^j(\Mat{x}) ,
    \end{align}
    \end{subequations}
where $\Mat{x}:=(x^0,\dots,x^3)=u(x)\in\field[R]$ are the coordinates of
$x\in\base$ relative to $u$ and similarly for $\Mat{x}'$. In particular, we
have to put in the above formulae
    \begin{equation}    \label{3.15}
A^{-1}(x) = I(u\mapsto u';x)
    \end{equation}
in a case when the change $u\mapsto u'$ induces $\{e^i\}\mapsto\{e^{\prime\,
i}\}$.

	Since the arguments leading to~\eref{2.8-2} are completely valid in the
framework of fibre bundle approach, we can claim the existence of an operator
$U$ on the system's space of states such that
    \begin{equation}    \label{3.17}
\varphi'_i(x) = U\circ\varphi_i(x)\circ U^{-1}
    \end{equation}
or, equivalently,
    \begin{equation}    \label{3.18}
\varphi'_{u',i}(\Mat{x}) = U\circ\varphi_{u,i}(\Mat{x})\circ U^{-1} .
    \end{equation}
Consequently, now the analogues of~\eref{2.12} read
	\begin{align}	\label{3.19}
& \varphi'_i(x)
= U\circ\varphi_i(x)\circ U^{-1}
= (A^{-1}(x))_i^j \varphi_j(x)
\\		    \label{3.20}
& \varphi'_{u',i}(\Mat{x})
= U\circ\varphi_{u,i}(\Mat{x})\circ U^{-1}
= (A^{-1}(x))_i^j \varphi_{u,j}(\Mat{x}) .
	\end{align}

	It should be noted, in all cases, quantum or classical ones, the set of
admissible matrices $A(x)$ or $I(u\mapsto u';x)$ defining the changes
$\{e^i\}\mapsto\{e^{\prime\, i}\}$ generally does not coincide with the one of
all non-degenerate matrices. Said differently, the set of admissible frames
$\{e^i\}$ for a given field need not to be identical with the one of all
frames. For instance, the set of admissible frames for a vector field, which is
a section of the tangent bundle $(T(M),\pi_T,\base)$, which are naturally
associated with local coordinates, is the one of all coordinate frames
(see~\eref{5.2b} below. Other example is a spinor field for which the
matrix~\eref{3.15} has a quite special form (see~\eref{5.2c} below).~%
\footnote{~%
However, at the time being seems to be open the problem for finding a
particular representative of the admissible frames (which will determine the
remaining ones).%
}
What concerns scalar fields, for them the matrix $A(x)$ is the identity one,
$A(x)=\openone$.

\subsection{Example: Relativistic quantum mechanics}
    \label{Subsect5.1}

	Let $\base$ be the Minkowski spacetime. Suppose $(\base,u)$ is a
global chart of $\base$ such that in the associated to it coordinate system
$\{u^\mu\}$ the metric (tensor) has a Lorentzian form~
\footnote{~%
It is (often silently) accepted that the chart $(\base,\id_\base)$ and the
associated to it Cartesian coordinate system have the property just described.
This hypothesis should be included in the (physical) definition of the Minkowski
spacetime.%
}
i.e., in physical terms, $\{u^\mu\}$ is (a part of) an inertial reference
frame. The admissible (inertial) coordinate systems are obtained from one
another via Poincar\'e transformations~\eref{2.3-2}. So, if we set
$x^\mu=u^\mu(x)$ for some $x\in\base$ and $\Mat{x}:=(x^0,\dots,x^3)^\top$, then
an admissible coordinate change can be represented as
    \begin{equation}    \label{5.0}
\Mat{x}\mapsto \Mat{x}'=\Lambda\Mat{x}+a ,
    \end{equation}
where $\Lambda$ is a $4\times 4$
matrix representing 4\ndash rotation and $a=(a^0,\dots,a^3)^\top$ is the
collection of the components  of a 4\ndash vector representing a 4\ndash
translation.

	We associate a matrix $I(u\mapsto u';x)$ (see~\eref{3.7}) with the
change~\eref{5.0} such that
    \begin{equation}    \label{5.1}
I(u\mapsto u';x) = I(\Lambda,x)
    \end{equation}
depending on $\Lambda$, where the mapping $\Lambda\mapsto I(\Lambda,x)$ is a
representation of the Poincar\'e group, and $x\in\base$; the
independence of $a$ reflects the translational invariance of the theory.~%
\footnote{~%
In a case of local transformations, the matrix $\Lambda$ may depend on the
point $x\in\base$.%
}
The particular choice of the mapping $\Lambda\mapsto I(\Lambda,x)$ characterizes
the particular field under considerations.~%
\footnote{~%
It also depends on the class of inertial reference frames as a whole. However,
this class is fixed in our case via the requirement the coordinate system
associated with the global chart $(\base,\id_\base)$ to be (a part of) an
inertial reference frame.%
}
For example, we have:
    \begin{subequations}    \label{5.2}
    \begin{alignat}{2}
			    \label{5.2a}
& I(\Lambda,x) = \openone
&\qquad &\text{for spin-0 (scalar) field}
\\			    \label{5.2b}
& (I(\Lambda,x))_\mu^\nu
= \frac{\pd u^{\prime\,\nu}}{\pd u^\mu}\Big|_x
=\Sprindex[\Lambda]{\mu}{\nu}
&\qquad &\text{for spin-1 (vector) field}
\\			    \label{5.2c}
& I(\Lambda,x)
= \exp{\bigl( -\frac{\iu}{4}\omega \sigma_{\mu\nu} I_{n}^{\mu\nu} \bigr) }
&\qquad &\text{for spin-{\footnotesize $\frac{1}{2}$} (spinor) field} ,
    \end{alignat}
    \end{subequations}
where $\iu$ is the imaginary unit,
$\sigma_{\mu\nu}=\frac{1}{2}[\gamma_{\mu},\gamma_{\nu}]$ is the commutator of
the Dirac matrices $\gamma_{\mu}$, $\Lambda$ is a 4\ndash rotation at an angle
$\omega$ around an axis $n$, and $I_n$ is the generator of this rotation.

	Substituting~\eref{5.1} into~\eref{3.9'}, we get
    \begin{gather}    \label{5.3}
\varphi_{u'}^{\prime\, i} (\Lambda\Mat{x}+a)
= I_j^i(\Lambda,x) \varphi_u^j(\Mat{x})
\\\intertext{or equivalently
(see~\eref{2.3-4a})}
		\tag{\ref{5.3}$^\prime$}
		    \label{5.3'}
\varphi_{u'}^{\prime\, i} (\Mat{x})
= I_j^i(\Lambda,x) \varphi_u^j(\Lambda^{-1}(\Mat{x}-a)) .
    \end{gather}

\subsection{Example: Quantum field theory}
    \label{Subsect5.2}

	In quantum field theory the field components $\varphi_i$ are known as
field operators and are suppose to be operators (precisely, operator-valued
distributions) acting on the Hilbert space of the system. In this case there
exists a representation $\mathcal{U}$ of the Poincar\'e group on the space of
the stated vectors such that a coordinate change
$\Mat{x}\mapsto\Mat{x}'=\Lambda\Mat{x}+a$ induces the transformation
(cf.~\eref{2.8-2})
    \begin{equation}    \label{5.4}
\varphi_{u, i}(\Mat{x})\mapsto
\varphi'_{u', i}(\Mat{x})
=
\mathcal{U}(\Lambda,a)\circ\varphi_{u , i}(\Mat{x})
\circ \mathcal{U}^{-1}(\Lambda,a) .
    \end{equation}

	Thus, similarly to~\eref{2.17}--\eref{2.19}, we can write
    \begin{align}    \label{5.5}
& \varphi_{u',i}^{\prime} (\Mat{x})
= I_i^j(\Lambda,x) \varphi_{u,j}(\Lambda^{-1}(\Mat{x}-a))
\\		    \label{5.6}
& \varphi'_{u', i}(\Mat{x})
=
\mathcal{U}(\Lambda,a)\circ\varphi_{u , i}(\Mat{x})
\circ \mathcal{U}^{-1}(\Lambda,a)
\\		    \label{5.7}
& \mathcal{U}(\Lambda,a)\circ\varphi_{u, i}(\Mat{x})
\circ \mathcal{U}^{-1}(\Lambda,a)
=
I_i^j( \Lambda;x ) \varphi_{u,j}( \Lambda^{-1}(\Mat{x}-a) )
    \end{align}
which can be combined into
    \begin{equation}    \label{5.8}
\varphi'_{u', i}(\Mat{x})
=
\mathcal{U}(\Lambda,a)\circ\varphi_{u, i}(\Mat{x})
\circ \mathcal{U}^{-1}(\Lambda,a)
=
I_i^j( \Lambda;x ) \varphi_{u,j}( \Lambda^{-1}(\Mat{x}-a) )
    \end{equation}
where the particular form of the matrix $[I_i^j( \Lambda;x)]$ depends on the
particular field under considerations (see, e.g.,~\eref{5.2}).

	In terms of the components $\varphi_i$ of the section $\varphi$
(see~\eref{3.10} and~\eref{3.13-1}) the relations~\eref{5.5}--\eref{5.7} read
    \begin{align}    \label{5.9}
\varphi'_i(x)
&= I_i^j(\Lambda,x) \varphi_j(x)
\\		    \label{5.10}
\varphi'_i(x)
&= \mathcal{U}(\Lambda,a)\circ \varphi(x) \circ \mathcal{U}^{-1}(\Lambda,a)
\\		    \label{5.11}
\mathcal{U}(\Lambda,a)\circ \varphi(x) \circ \mathcal{U}^{-1}(\Lambda,a) .
&=
I_i^j(\Lambda,x) \varphi_j(x) .
    \end{align}
According to~\eref{3.14b}, \eref{3.15} and~\eref{5.1}, the test functions
transform into
    \begin{equation}    \label{5.12}
f^{\prime\, i}_{u'}(\Lambda\Mat{x}+a)
=
(I^{-1}(\Lambda,x))_j^i f_u^j(\Mat{x})
    \end{equation}
or equivalently
    \begin{equation}    \label{5.12'}
		\tag{\ref{5.12}$^\prime$}
f^{\prime\, i}_{u'}(\Mat{x})
=
(I^{-1}(\Lambda,x))_j^i f_u^j(\Lambda^{-1}(\Mat{x}-a))
    \end{equation}
which can also be written as (see~\eref{3.13-1}, cf.~~\eref{3.13c})
    \begin{equation}    \label{5.13}
f^{\prime\, i}(\Mat{x})
=
(I^{-1}(\Lambda,x))_j^i f^j(x) .
    \end{equation}

	In this way, we see that in the fibre bundle aproach both the classical
and quantum fields transform according to passive coordinate transformations
when one changes the frames of reference contrary to the conclusions at the
end of section~\ref{Sect3} in which fibre bundles were not involved.

	We  also observed that in the bundle approach for the derivation of
equations like~\eref{5.5} and~\eref{5.12'} one needs not to make additional
hypotheses in contrast to the derivation of equalities like~\eref{2.9}
and~\eref{2.9-1}.


\section {Heisenberg relations}
\label{Sect5}

	As Heisenberg relations or equations in quantum field theory are known a
kind of commutation relations between the field operators and the generators (of
a representation) of a group acting on system's Hilbert space of states. Their
(global) origin is in the equations~\eref{2.8-2} (or,
equivalently,~\eref{2.8-1} or~\eref{2.6} and~\eref{2.8}) in which it is
expected that the transformed field operators  $\varphi'_i$ can be expressed
explicitly by means of $\varphi_i$ via equations like~\eref{2.9}. If the
elements $U$ (of the representation) of the group are labeled by
$b=(b^1,\dots,b^s)\in\field^s$ for some $s\in\field[N]$ (we are dealing, in
fact, with a Lie group), \ie we may write $U(b)$ for $U$, then the
corresponding Heisenberg relations are obtained from~\eref{2.8-2} with $U(b)$
for $U$ by differentiating it with respect to $b^\omega$, $\omega=1,\dots,s$,
and then setting $b=b_0$, where $b_0\in\field^s$ is such that $U(b_0)$ is the
identity element.

	The above shows that the Heisenberg relations are from pure
geometric-group-theoretical origin and the only physics in them is the
motivation leading to~\eref{2.6}. However, there are strong evidences that to
the Heisenberg relations can be given dynamical/physical sense by
identifying/replacing in them the generators (of the representation) of the
group by the corresponding operators of conserved physical quantities if the
system considered is invariant with respect to this group (see, \eg the
discussion in~\cite[\S~68]{Bjorken&Drell-2}).

	In subsections~\ref{Subsect6.1}-\ref{Subsect6.3}, we consider
Heisenberg relations in the non-bundle approach (see sections~\ref{Sect2}
and~\ref{Sect3}), while in subsection~\ref{Subsect6.4} they are investigated
on the ground of fibre bundles.

\subsection{The Poincar\'e group}
\label{Subsect6.1}

	Suppose we study a quantum field with components $\varphi_i$ relative
to two reference frames connected by a general Poincar\'e
transformation~\eref{2.3-2}. Then the ``global' version of the Heisenberg
relations is expressed by the second equality in~\eref{2.12}, \ie
	\begin{equation}	\label{6.1}
U(\Lambda,a)\circ\varphi_i(x)\circ U^{-1}(\Lambda,a)
=
{D}_i^j(\Lambda,a)\varphi_j(\Lambda x+a) ,
	\end{equation}
where $U$ (resp.\ $D$) is a representation of the Poincar\'e group on the space
of state vectors (resp.\ on the space of field operators),
 $U(\Lambda,a)$ (resp.\ $\Mat{D}(\Lambda,a)=[{D}_i^j(\Lambda,a)]$) is the
mapping (resp.\ the matrix of the mapping) corresponding via $U$ (resp.\ $D$)
to~\eref{2.3-2}, the index $u$ in $\varphi_{u,i}$ is omitted (\ie we write
$\varphi_i$ for $\varphi_{u,i}$) and the point $x\in\base$ is identified with
$\Mat{x}=u(x)\in\field[R]^4$. Since for $\Lambda=\openone$ and
$a=\Mat{0}\in\field[R]^4$ is fulfilled $u'(x)=u(x)$, we have
    \begin{equation}    \label{6.2}
U(\openone,\Mat{0}) = \id
\qquad
\Mat{D}(\openone,\Mat{0}) = \openone ,
    \end{equation}
where $\id$ is the corresponding identity mapping and $\openone$ stands for the
corresponding identity matrix. Let $\Lambda=[\Sprindex[\Lambda]{\nu}{\mu}]$,
 $\Lambda^{\mu\nu}:=\eta^{\nu\lambda}\Sprindex[\Lambda]{\lambda}{\mu}$, with
$\eta^{\mu\nu}$ being the components of the Lorentzian metric with signature
$(-+++)$, and define
    \begin{subequations}    \label{6.3}
    \begin{align}    \label{6.3a}
T_\mu
& := \frac{\pd U(\Lambda,a)}{\pd a^\mu}
\Big|_{(\Lambda,a)=(\openone,\Mat{0})}
\\		    \label{6.3b}
S_{\mu\nu}
& := \frac{\pd U(\Lambda,a)}{\pd \Lambda^{\mu\nu}}
\Big|_{(\Lambda,a)=(\openone,\Mat{0})}
\\		    \label{6.3c}
H_{j\mu}^{i}
& := \frac{\pd {D}_j^i(\Lambda,a)}{\pd a^\mu}
\Big|_{(\Lambda,a)=(\openone,\Mat{0})}
\\		    \label{6.3d}
I_{j\mu\nu}^{i}
& := \frac{\pd {D}_j^i(\Lambda,a)}{\pd \Lambda^{\mu\nu}}
\Big|_{(\Lambda,a)=(\openone,\Mat{0})} .
    \end{align}
    \end{subequations}
The particular form of the numbers $I_{j\mu\nu}^i$ depends on the field under
consideration. In particular, we have (see~\eref{5.2})
    \begin{subequations}    \label{6.5}
    \begin{alignat}{2}
			    \label{6.5a}
& I_{1\mu\nu}^1 = 0
&\qquad &\text{for spin-0 (scalar) field}
\\			    \label{6.5b}
& I_{\rho\mu\nu}^\sigma
=
\delta_\mu^\sigma \eta_{\nu\rho} - \delta_\nu^\sigma \eta_{\mu\rho}
&\qquad &\text{for spin-1 (vector) field}
\\			    \label{6.5c}
& [ I_{j\mu\nu}^i ]_{i,j=1}^4
=
-\frac{1}{2} \iu \sigma_{\mu\nu}
&\qquad &\text{for spin-{\footnotesize $\frac{1}{2}$} (spinor) field} .
    \end{alignat}
    \end{subequations}

	Differentiating~\eref{6.1} relative to $a^\mu$ and setting after that
$(\Lambda,a)=(\openone,\Mat{0})$, we find
    \begin{equation}    \label{6.4}
[T_\mu,\varphi_i(x)]_{\_} = \pd_\mu\varphi_i(x) + H_{i\mu}^{j}\varphi_j(x) ,
    \end{equation}
where $[A,B]_{\_}:=AB-BA$ is the commutator of some operators or matrices $A$
and $B$.
	Since the field theories considered at the time being are invariant
relative to spacetime translation of the coordinates, \ie with respect to
$\Mat{x}\mapsto\Mat{x}+a$, further we shall suppose that (cf.~\eref{6.3c})
    \begin{equation}    \label{6.6}
H_{j\mu}^{i} = 0.
    \end{equation}
In this case~\eref{6.4} reduces to
    \begin{subequations}    \label{6.7}
    \begin{equation}    \label{6.7a}
[T_\mu,\varphi_i(x)]_{\_} = \pd_\mu\varphi_i(x).
    \end{equation}
Similarly, differentiation~\eref{6.1} with respect to
$\Lambda^{\mu\nu}$ and putting after that $(\Lambda,a)=(\openone,\Mat{0})$, we
obtain
    \begin{equation}    \label{6.7b}
[S_{\mu\nu},\varphi_i(x)]_{\_}
= x_\mu \pd_\nu\varphi_i(x) - x_\nu \pd_\mu\varphi_i(x) +
I_{i\mu\nu}^{j}\varphi_j(x)
    \end{equation}
    \end{subequations}
where $x_\mu:=\eta_{\mu\nu}x^\nu$. The equations~\eref{6.7} are
identical up to notation with~\cite[eqs.(11.70) and~(11.73)]{Bjorken&Drell-2}.
Note that for complete correctness one should write $\varphi_{u,i}(\Mat{x})$
instead of $\varphi_i(x)$ in~\eref{6.7}, but we do not do this to keep our
results near to the ones accepted in the physical
literature~\cite{Roman-QFT,Bjorken&Drell,Bogolyubov&Shirkov}.

	As we have mentioned earlier, the particular Heisenberg
relations~\eref{6.7} are from pure geometrical-group-theoretical origin. The
following heuristic remark can give a dynamical sense to them. Recalling that
the translation (resp.\ rotation) invariance of a (Lagrangian) field theory
results in the conservation of system's momentum (resp.\ angular momentum)
operator $P_\mu$ (resp.\ $M_{\mu\nu}$) and the correspondences
    \begin{equation}    \label{6.9}
\ih T_{\mu} \mapsto P_{\mu}
\qquad
\ih S_{\mu\nu} \mapsto M_{\mu\nu} ,
    \end{equation}
with $\hbar$ being the Planck's constant (divided by $2\pi$), one may suppose
the validity of the Heisenberg relations
    \begin{subequations}    \label{6.10}
    \begin{align}    \label{6.10a}
[P_\mu,\varphi_i(x)]_{\_}
&= \ih\pd_\mu\varphi_i(x)
\\		    \label{6.10b}
[M_{\mu\nu},\varphi_i(x)]_{\_}
&= \ih \{ x_\mu \pd_\nu\varphi_i(x) - x_\nu \pd_\mu\varphi_i(x) +
I_{i\mu\nu}^{j}\varphi_j(x) \} .
    \end{align}
    \end{subequations}
However, one should be careful when applying the last two equations  in the
Lagrangian formalism as they are external to it and need a particular proof in
this approach; \eg they hold in the free field
theory~\cite{Bjorken&Drell,bp-MP-book}, but a general proof seems to be missing.
In the axiomatic quantum field
theory~\cite{Roman-QFT,Bogolyubov&et_al.-AxQFT,Bogolyubov&et_al.-QFT} these
equations are identically valid as in it the generators of the translations
(rotations) are identified up to a constant factor with the components of the
(angular) momentum operator, $P_\mu=\ih T_\mu$ ($M_{\mu\nu}=\ih S_{\mu\nu}$).


\subsection{Internal transformations}
\label{Subsect6.2}

	In our context, an internal transformation is a change of the reference
frame such that the spacetime coordinates remain unchanged.

	Let $G$ be a group whose elements $g_b$ are labeled by $b\in\field^s$
for some $s\in\field[N]$.~%
\footnote{~%
In fact, we are dealing with an $s$-dimensional Lie group and $b\in\field^s$ are
the (local) coordinates of $g_b$ in some chart on $G$ containing $g_b$ in its
domain.%
}
Consider two reference frames $(u,\{e^i\})$ and$(u,\{e'^i\})$ (see
section~\ref{Sect2}) connected via~\eref{2.2-1} with
    \begin{equation}    \label{6.12}
A^{-1}(x) = I(b)
    \end{equation}
where $I \colon G\mapsto \GL(\dim V,\field)$ is a matrix representation of $G$
and $I \colon G\ni g_b\mapsto I(b)\in \GL(\dim V,\field)$. Notice, here and
below we label the elements of a frame with \emph{superscripts} instead
with subscripts, as in section~\ref{Sect2}. As a result of~\eref{2.8-2},
\eref{2.6-1} and~\eref{2.6-2} the field operators transform into
    \begin{equation}    \label{6.13}
\varphi'_{u,i}(r) = U(b)\circ\varphi_{u,i}(r)\circ U^{-1}(b)
    \end{equation}
where $U$ is a representation of $G$ on the Hilbert space of state vectors and
$U \colon G\ni g_b\mapsto U(b)$. Thus~\eref{2.19} now reduces to
    \begin{equation}    \label{6.14}
U(b)\circ\varphi_{u,i}(r)\circ U^{-1}(b) = I_i^j(b) \varphi_{u,j} (r)
    \end{equation}
due to $u'=u$ in the case of internal trasformations considered here.

	Suppose $b_0\in\field^s$ is such that $g_{b_0}$ is the identity element
of $G$ and define
    \begin{equation}    \label{6.14-1}
Q_\omega : = \frac{\pd U(b)}{\pd b^\omega} \Big|_{b=b_0}
\qquad
I_{i \omega}^j : = \frac{\pd I_i^j(b)}{\pd b^\omega} \Big|_{b=b_0}
    \end{equation}
where $b=(b^1,\dots,b^s)$ and $\omega=1,\dots,s$. Then,
differentiating~\eref{6.14} with respect to $b^\omega$ and putting in the
result $b=b_0$, we get the following Heisenberg relation
    \begin{equation}    \label{6.15}
[ Q_\omega, \varphi_{u,i}(r) ]_{\_} = I_{i\omega}^j \varphi_{u,j}(r)
    \end{equation}
or, if we identify $x\in\base$ with $r=u(x)$ and omit the subscript $u,$
    \begin{equation}    \label{6.16}
[ Q_\omega, \varphi_{i}(x) ]_{\_} = I_{i\omega}^j \varphi_{j}(x) .
    \end{equation}

	To make the situation more familiar, consider the case of one\ndash
dimensional group $G$, $s=1$, when $\omega=1$ due to which we shall identify
$b^1$ with $b=(b^1)$. Besides, let us suppose that
    \begin{equation}    \label{6.17}
I(b) = \openone \exp(f(b)-f(b_0))
    \end{equation}
for some $C^1$ function $f$. Then~\eref{6.16} reduces to
    \begin{equation}    \label{6.18}
[ Q_1, \varphi_{i}(x) ]_{\_} = f'(b_0) \varphi_i(x) ,
    \end{equation}
where $f'(b):=\frac{\od f(b)}{\od b}$. In particular, if we are dealing with
phase transformations, \ie
    \begin{equation}    \label{6.19}
U(b) = \e^{\frac{1}{\iu e} b Q_1}
\quad
I(b) = \openone \e^{-\frac{q}{\iu e} b}
\qquad
b\in\field[R]
    \end{equation}
for some constants $q$ and $e$ (having a meaning of charge and unit charge,
respectively) and operator $Q_1$ on system's Hilbert space of states (having a
meaning of a charge operator), then~\eref{2.11} and~\eref{6.18} take the
familiar form~\cite[eqs.~(2.81) and~(2-80)]{Roman-QFT}
    \begin{align}    \label{6.20}
& \varphi'_i(x)
=
\e^{\frac{1}{\iu e} b Q_1}\circ \varphi_i(x) \circ \e^{-\frac{1}{\iu e} b Q_1}
=
\e^{-\frac{q}{\iu e} b} \varphi(x)
\\		    \label{6.21}
& [Q_1,\varphi_i(x)]_{\_} = -q \varphi_i(x) .
    \end{align}

	The considerations in the framework of Lagrangian formalism invariant
under  phase
transformations~\cite{Roman-QFT,Bjorken&Drell,Bogolyubov&Shirkov} implies
conservation of the charge operator $Q$ and suggests the correspondence
(cf.~\eref{6.9})
    \begin{equation}    \label{6.22}
Q_1 \mapsto Q
    \end{equation}
which in turn suggests the Heisenberg relation
    \begin{equation}    \label{6.23}
[Q,\varphi_i(x)]_{\_} = - q \varphi_i(x) .
    \end{equation}
We should note that this equation is external to the Lagrangian formalism and
requires a proof in it~\cite{bp-MP-book}.


\subsection{The general case}
\label{Subsect6.3}

	As we saw in the previous subsections, the corner stone of the (global)
Heisenberg relations is the equation~\eref{2.19},
    \begin{equation}    \label{6.24}
 U\circ\varphi_{u,i}(r)\circ U^{-1}
=
\frac{\pd( u '\circ u ^{-1})(r)}{\pd r}
	\bigl( A^{-1}( u ^{-1}(r)) \bigr)_i^j
	\varphi_{u,j}( ( u '\circ u ^{-1})(r) ) .
    \end{equation}
Now, following the ideas at the beginning of this section, we shall demonstrate
how from it can be derived Heisenberg relations in the general case.

	Let $G$ be an $s$-dimensional, $s\in\field[N]$, Lie group. Without going
into details, we admit that its elements are labeled by
$b=(b^1,\dots,b^s)\in\field^s$ and $g_{b_0}$ is the identity element of $G$ for
some fixed $b_0\in\field^s$. Suppose that there are given three representations
$H$, $I$ and $U$ of $G$ and consider frames of reference with the following
properties:
    \begin{enumerate}
\item
\(
H\colon G\ni g_b\mapsto H_b
\colon \field[R]^{\dim \base}\to \field[R]^{\dim \base}
\) and any change $(U,u)\mapsto(U',u')$ of the charts of $M$ is such
that $u'\circ u^{-1}=H_b$ for some $b\in\field^s$.

\item
$I \colon G\ni g_b\mapsto I(b)\in\GL(\dim V,\field)$
and any change $\{e^i\}\mapsto\{e^{\prime\mspace{2mu} i}=A^i_je^j\}$ of the
frames in $V$ is such that $A^{-1}(x)=I(b)$ for all $x\in\base$ and some
$b\in\field^s$.

\item
 $U \colon G\ni g_b\mapsto U(b)$, where $U(b)$ is an operator on the space of
state vectors, and a change $(u,\{e^i\})\mapsto(u',\{e'^i\})$ of the
reference frame entails~\eref{2.8-2} with $U(b)$ for $U$.
    \end{enumerate}

	Under the above hypotheses equation~\eref{6.24} transforms into
    \begin{equation}    \label{6.25}
U(b)\circ\varphi_{u,i}(r)\circ U^{-1}(b)
=
\det\Bigl[ \frac{\pd (H_b(r))^k}{\pd r^l} \Bigr]
I_i^j(b) \varphi_{u,j}(H_b(r))
    \end{equation}
which can be called \emph{global Heisenberg relation} in the particular
situation. The next step is to differentiate this equation with respect to
$b^\omega$, $\omega=1,\dots,s$, and then to put  $b=b_0$ in the result. In this
way we obtain the following \emph{(local) Heisenberg relation}
    \begin{equation}    \label{6.26}
[U_\omega,\varphi_{u,i}(r)]_{\_}
=
  \Delta_\omega(r)\varphi_{u,i}(r)
+ I_{i\omega}^j \varphi_{u,j}(r)
+ (h_\omega(r))^k \frac{\pd \varphi_{u,i}(r)}{\pd r^k} ,
    \end{equation}
where
    \begin{subequations}    \label{6.27}
    \begin{align}    \label{6.27a}
U_\omega
& := \frac{\pd U(b)}{\pd b^\omega} \Big|_{b=b_0}
\\		    \label{6.27b}
\Delta_\omega(r)
& :=
\left.
\frac{\pd \det\Bigl[ \frac{\pd (H_b(r))^i}{\pd r^j} \Bigr] } {\pd b^\omega}
\right|_{b=b_0}
\in  \field[K]
\\		    \label{6.27c}
I_{i\omega}^j
& :=
\frac{\pd I_i^j(b)}{\pd b^\omega} \Big|_{b=b_0} \in\field
\\		    \label{6.27d}
h_\omega
& :=
\frac{\pd H_b}{\pd b^\omega} \Big|_{b=b_0}
 \colon \field[R]^{\dim \base}\to \field[R]^{\dim \base} .
    \end{align}
    \end{subequations}

	In particular, if $H_b$ is linear and non-homogeneous, \ie
 $H_b(r)=H(b)\cdot r + a(b)$ for some $H(b)\in\GL(\dim\base,\field[R])$ and
$a(b)\in\field^{\dim\base}$ with $H(b_0)=\openone$ and $a(b_0)=\Mat{0}$, then
($\Tr$ means trace of a matrix or operator)
    \begin{equation}    \label{6.28}
\Delta_\omega(r)
= \frac{\pd \det(H(b))}{\pd b^\omega} \Big|_{b=b_0}
= \frac{\pd \Tr(H(b))}{\pd b^\omega} \Big|_{b=b_0}
\qquad
h_\omega(\,\cdot\,)
= \frac{\pd H(b)}{\pd b^\omega} \Big|_{b=b_0} \cdot  (\,\cdot\,)
+ \frac{\pd a(b)}{\pd b^\omega} \Big|_{b=b_0}
    \end{equation}
as $\frac{\pd \det B}{\pd b_i^j}\big|_{B=\openone}=\delta_j^i$ for any square
matrix $B=[b_i^j]$. In this setting the Heisenberg relations corresponding to
Poincar\'e transformations (see subsection~\ref{Subsect6.1}) are described by
$b\mapsto(\Lambda^{\mu\nu},a^\lambda)$, $H(b)\mapsto\Lambda$, $a(b)\mapsto a$
and $I(b)\mapsto I(\Lambda)$, so that
 $U_\omega\mapsto(S_{\mu\nu},T_\lambda)$,
 $\Delta_\omega(r)\equiv 0$,
$I_{i\omega}^j\mapsto(I_{i \mu\nu}^j,0)$ and
\(
(h_\omega(r))^k\frac{\pd }{\pd r^k}
\mapsto
r_\mu\frac{\pd }{\pd r^\nu} - r_\nu\frac{\pd }{\pd r^\mu} .
\)

	The case of internal transformations, considered in the previous
subsection, corresponds to $H_b=\id_{\field[R]^{\dim\base}}$ and, consequently,
in it $\Delta_\omega(r)\equiv 0$ and $h_\omega=0$.

\subsection{Fibre bundle approach}
\label{Subsect6.4}

	The origin of the Heisenberg relations on the background of fibre
bundle setting is in any one of the equivalent second equalities in~\eref{3.19}
or~\eref{3.20},
	\begin{align}	\label{6.29}
U\circ\varphi_i(x)\circ U^{-1}
&= (A^{-1})_i^j(x) \varphi_j(x)
\\			\tag{\ref{6.29}$'$}
			\label{6.29'}
U\circ\varphi_{u,i}(\Mat{x})\circ U^{-1}
&= (A^{-1})_i^j(x) \varphi_{u,j}(\Mat{x}) .
	\end{align}

	Similarly to subsection~\ref{Subsect6.3}, consider a Lie group $G$, its
representations $I$ and $U$ and reference frames with the following properties:
    \begin{enumerate}
\item
$I \colon G\ni g_b\mapsto I(b)\in\GL(\dim V,\field)$
and the changes $\{e_i\}\mapsto\{e'^i=A_j^ie^j\}$ of the frames in $V$ are such
that $A^{-1}(x)=I(b)$ for all $x\in\base$ and some $b\in\field^s$.

\item
 $U \colon g\ni g_b\mapsto U(b)$, where $U(b)$ is an operator on the space of
state vectors, and the changes $(u,\{e^i\})\mapsto(u',\{e'^i\})$ of the
reference frames entail~\eref{2.8-2} with $U(b)$ for $U$.
    \end{enumerate}

    \begin{Rem}    \label{Rem6.1}
One can consider also simultaneous coordinate changes
$u\mapsto u'=H_b\circ u$ induced by a representation
\(
H \colon G\ni g_b\mapsto H_b
   \colon \field[R]^{\dim\base}\to \field[R]^{\dim\base} ,
\)
as in subsection~\ref{Subsect6.3}. However such a supposition does not influence
our results as the basic equations~\eref{6.30} and~\eref{6.30'} below are
independent from it; in fact, equation~\eref{6.30} is coordinate-independent,
while~\eref{6.30'} is its version valid in any local chart $(U,u)$ as
$\varphi_{u}:=\varphi\circ u^{-1}$ and $\Mat{x}:=u(x)$.

    \end{Rem}

Thus equations~\eref{6.29} and~\eref{6.29'} transform into the following
\emph{global Heisenberg relations} (cf.~\eref{6.25})
	\begin{align}	\label{6.30}
U(b)\circ\varphi_i(x)\circ U^{-1}(b)
&= I_i^j(b) \varphi_j(x)
\\			\tag{\ref{6.30}$'$}
			\label{6.30'}
U(b)\circ\varphi_{u,i}(\Mat{x})\circ U^{-1}(b)
&= I_i^j(b) \varphi_{u,j}(\Mat{x}) .
	\end{align}
Differentiating~\eref{6.30} with respect to $b^\omega$ and then putting
$b=b_0$, we derive the following \emph{(local) Heisenberg relations}
	\begin{align}	\label{6.31}
[U_\omega,\varphi_i(x)]_{\_}
&= I_{i\omega}^j \varphi_j(x)
\\\intertext{or its equivalent version (cf.~\eref{6.26})}
			\tag{\ref{6.31}$'$}
			\label{6.31'}
[U_\omega,\varphi_{u,i}(\Mat{x})]_{\_}
&= I_{i\omega}^j \varphi_{u,j}(\Mat{x}) ,
	\end{align}
where
    \begin{subequations}    \label{6.32}
    \begin{align}    \label{6.32a}
U_\omega
& := \frac{\pd U(b)}{\pd b^\omega} \Big|_{b=b_0}
\\		    \label{6.32b}
I_{i\omega}^j
& :=
\frac{\pd I_i^j(b)}{\pd b^\omega} \Big|_{b=b_0} .
    \end{align}
    \end{subequations}
Recalling~\eref{3.10}, we can rewire the Heisenberg relations obtained as
    \begin{equation}    \label{6.33}
[U_\omega,\varphi]_{\_} = I_{i\omega}^{j} \varphi_j e^i .
    \end{equation}
One can prove that the r.h.s.\ of this equation is independent of the
particular frame $\{e^i\}$ in which it is represented.

	The case of Poncar\'e transformations is described by the replacements
$b\mapsto(\Lambda^{\mu\nu},a^\lambda)$, $U_\omega\mapsto(S_{\mu\nu},T_\lambda)$
and $I_{i\omega}^{j}\mapsto(I_{i\mu\nu}^j,0)$ (for translation invariant
theory) and, consequently, the equations~\eref{6.30} and~\eref{6.30'} now read
	\begin{align}	\label{6.33-1}
U(\Lambda,a)\circ\varphi_i(x)\circ U^{-1}(\Lambda,a)
&= I_i^j(\Lambda,a) \varphi_j(x)
\\			\tag{\ref{6.33-1}$'$}
			\label{6.30-1'}
U(\Lambda,a)\circ\varphi_{u,i}(\Mat{x})\circ U^{-1}(\Lambda,a)
&= I_i^j(\Lambda,a) \varphi_{u,j}(\Mat{x}) .
	\end{align}
Hence, for instance, the Heisenberg relation~\eref{6.31} now takes the form
(cf.~\eref{6.7})
    \begin{subequations}    \label{6.34}
    \begin{align}    \label{6.34a}
[T_{\mu},\varphi_i(x)]_{\_} &= 0
\\		    \label{6.34b}
[S_{\mu\nu},\varphi_i(x)]_{\_} &= I_{i\mu\nu}^{j} \varphi_j(x) .
    \end{align}
    \end{subequations}
Respectively, the correspondences~\eref{6.9} transform these equations into
    \begin{subequations}    \label{6.35}
    \begin{align}    \label{6.35a}
[P_{\mu},\varphi_i(x)]_{\_} &= 0
\\		    \label{6.35b}
[M_{\mu\nu},\varphi_i(x)]_{\_} &= I_{i\mu\nu}^{j} \varphi_j(x)
    \end{align}
    \end{subequations}
which now replace~\eref{6.10}.

	Since equation~\eref{6.10a} (and partially equation~\eref{6.10b}) is
(are) the corner stone for the particle interpretation of quantum field
theory~\cite{Bogolyubov&Shirkov,Bjorken&Drell,bp-MP-book}, the
equation~\eref{6.35a} (and partially equation~\eref{6.35b}) is (are) physically
unacceptable if one wants to retain the particle interpretation in the fibre
bundle approach to the theory. For this reason, it seems that the
correspondences~\eref{6.9} should \emph{not} be accepted in the fibre bundle
approach to quantum field theory, in which~\eref{6.7} transform
into~\eref{6.34}. However, for retaining the particle interpretation one can
impose~\eref{6.10} as subsidiary restrictions on the theory in the fibre bundle
approach. It is almost evident that this is possible if the frames used are
connected by linear homogeneous transformations with spacetime constant
matrices, $A(x)=\const$ or $\pd_\mu A(x)=0$. Consequently, if one wants to
retain the particle interpretation of the theory, one should suppose the
validity of~\eref{6.10} in some frame and, then, it will hold in the whole
class of frames obtained from the chosen one by transformations with
spacetime independent matrices.

	Since the general setting investigated above is independent of any
(local) coordinates, it describes also the fibre bundle version of the case of
internal transformations considered in subsection~\ref{Subsect6.2}. This
explains why equations line~\eref{6.15} and~\eref{6.31'} are identical but the
meaning of the quantities $\varphi_{u,i}$ and $I_{i\omega}^{j}$ in them is
different.~%
\footnote{~%
Note, now $I(b)$ is the matrix defining transformations of frames in the bundle
space, while in~\eref{6.19} it serves a similar role for frames in the vector
space $V$.%
}
In particular, in the case of phase transformations
    \begin{equation}    \label{6.36}
U(b) = \e^{\frac{1}{\iu e} b Q_1}
\quad
I(b) = \openone \e^{-\frac{q}{\iu e} b}
\qquad
b\in\field[R]
    \end{equation}
the Heisenberg relations~\eref{6.31} reduce to
    \begin{equation}    \label{6.37}
[Q_1,\varphi_i(x)]_{\_} = -q \varphi_i(x),
    \end{equation}
which is identical with~\eref{6.21}, but now $\varphi_i$ are the components of
the section $\varphi$ in $\{e^i\}$. The invariant form of the last relations is
    \begin{equation}    \label{6.38}
[Q_1,\varphi]_{\_} = -q \varphi
    \end{equation}
which is also a consequence from~\eref{6.33} and~\eref{6.20}.



\section {Conclusion}
\label{Conclusion}

	In this paper we have shown how the transformation laws of the
components of the classical and quantum fields arise.

 	We also have demonstrated how the Heisenberg equations can be derived
in the general case and in particular situations. They are from
pure geometrical origin and one should be careful when applying them in the
Lagrangian formalism in which they are subsidiary conditions, like the Lorenz
gauge in the quantum electrodynamics. In the general case they need not to be
consistent with the Lagrangian formalism and their validity should carefully be
checked. For instance, if one starts with field operators in the Lagrangian
formalism of free fields and adds to it the Heisenberg relations~\eref{6.10a}
concerning the momentum operator, then the arising scheme is not consistent as
in it start to appear distributions, like the Dirac delta function. This
conclusion leads to the consideration of the quantum fields as operator-valued
distribution in the Lagrangian formalism even for free fields. In the last
case, the Heisenberg relations concerning the momentum operator are consistent
with the Lagrangian formalism. Besides, they play an important role in the
particle interpretation of the so-arising theory.

	Let us now pay some attention on the observability and measurability of
the objects considered in this paper.

	We say that an object is observable if we can obtain information from
it directly or via its interaction with other objects(s) and we can detect (and
interpret) the result(s) of this interaction. It is measurable if we can
measure some its characteristics obtained in the process of observation by
assigning to them numerical values. One should assume that the physical fields
are observable and measurable as otherwise they cannot be studied rigorously.
At this point arises the question: are the components $\varphi_i$ of a field
$\varphi$ observable and possibly measurable?  Since the set $\{\varphi_i\}$ is
in a sense a projection of $\varphi$ on a reference frame, this question is
equivalent to the problem: are the reference frames observable and possibly
measurable?

	In the context of the present paper, a reference frame is a pair
$(u,e)$ of a coordinate system $u=(u^0,\dots,u^{\dim\base-1})$ on a manifold
$\base$, representing a spacetime model, and a frame $e=(e^1,\dots,e^n)$ in a
vector space $V$ or in the bundle space $E$ of a fibre bundle $(E,\pi,\base)$
with $n\in\field[N]$ being the dimension of $V$ or the fibre dimension of
$(E,\pi,\base)$.

	It is known that (locally) there exist experimental procedures which
(at least in principle) allow to be determined the coordinates
$\Mat{x}=(x^0,\dots,x^{\dim\base-1})$ of a spacetime point $x\in\base$ with
respect to some set of real physical objects; in fact, this set is the object
we mathematically describe via a coordinate system $u$ which, in this setting,
is defined by $u \colon x\mapsto u(x):=\Mat{x}$. Thus the coordinates of a
spacetime point are observable and measurable. From a coordinate system
$\{u^\mu\}$ on $\base$ are construct the coordinate frames
$\bigl\{ \frac{\pd }{\pd u^\mu} \bigr\}$ and  $\{\od u^\mu\}$ in respectively
the tangent and cotangent bundles over $\base$ and from them, by tensor
multiplication, can be constructed frames in the tensor bundles over $\base$.
Therefore we can claim that the coordinate frames in tensor bundles are
(indirectly) observable and measurable. Since the elements of a non-coordinate
frame in a tensor bundle can be represented as linear combinations (with
functions as coefficients) of the ones of an arbitrarily chosen coordinate
frame (in the overlap of their domains), we can also claim that the
non-coordinate frames in tensor bundles are also observable and measurable.
From here follows that if a tensor fieldd is observable and measurable, then
such are and its componets.

	Let us turn now our attention to elements of the (total) bundle space
of a fibre bundle, which is not a tensor bundle. Are the sections of such a
bundle and/or their components observable and, possible, measurable? In
particular, are the frames in non-tensor bundles observable and, possible,
measurable? It seems that the answer to these questions are in general
negative. For instance, the elements of a spinor bundle describing
(spin-$\frac{1}{2}$) Dirac field are not observable but from them can be
constructed observable quantities like the energy-momentum and charge
characteristics of the field~\cite{Bjorken&Drell}.

	At the time being there is only one phenomenon, known as the
Aharonov-Bohm effect~\cite{Aharonov&Bohm,Bernstein&Phillips,Baez&Muniain}, that
may lead to observability of elements of a non-tensor bundle. Its essence is that the electromagnetic potential
can give rise to directly observable results and, in this sense, are
observable. Let us suppose that this is true~%
\footnote{~%
There are some doubts in the reality of the Aharonov-Bohm effect.%
}
As the electromagnetic potentials $A_\mu$ are coefficients of a linear
connection $\nabla$ in  $C^1$ one\ndash dimensional vector bundle $(E,\pi,M)$,
we have $\nabla_{\frac{\pd }{\pd u^\mu}} e=A_\mu e$, where $\{u^\mu\}$ is a
coordinate system on $M$ and $\{e\}$ is a frame in $E$ consisting of a single
section $e \colon M\to E$ with non\ndash zero values. A change
 $(\{u^\mu\},e) \mapsto (\{u^{\prime \mu}\},e'=fe)$ with
$f \colon M\to \field[R]\setminus \{0\}$
implies~\cite[p.~356, eq.~(4.23)]{bp-NF-book}
\(
A_\mu\mapsto A'_\mu
=
\frac{\pd u^\nu}{\pd u'^\mu}
\bigl( A_\nu + \frac{\pd \ln f}{\pd u^\nu} \bigr) ;
\)
in particular, if $u'^\mu=u^\mu$, then we have a pure gauge transformation
$ A_\mu\mapsto A_\mu\mapsto A'_\mu = A_\mu + \frac{\pd \ln f}{\pd u^\mu}$.
Therefore the frame $\{e\}$ is observable via the electromagnetic potentials
$A_\mu$ in the reference frame $(\{u^\mu\},\{e\})$. However, since a change of
the frame $\{e\}$ is defined within a constant non\ndash vanishing factor,
which does not change $A_\mu$, the frame $\{e\}$ is not measurable regardless of
are the potentials $A_\mu$ measurable or not measurable.


%
%


\addcontentsline{toc}{section}{References}
\bibliography{Bozho-BiBTeX-Published-Works,Bozho-BiBTeX-References}
\bibliographystyle{unsrt}
\addcontentsline{toc}{subsubsection}{This article ends at page}



\end{document}